\documentclass[floats,floatfix, twocolumn,superscriptaddress,nofootinbib,showpacs]{revtex4-1}

\pdfoutput=1
\usepackage{amsfonts,amssymb,amsmath}
\usepackage[usenames,dvipsnames]{xcolor}
\usepackage[unicode,colorlinks=true,linkcolor=blue,urlcolor=blue,citecolor=gray]{hyperref}
\usepackage[all]{hypcap}
\usepackage{graphicx}
\usepackage{xspace}
\usepackage{tikz}
\usetikzlibrary{shapes,arrows}
\usetikzlibrary{matrix}
\usetikzlibrary{shapes.multipart}
\usetikzlibrary{er,positioning}
\usepackage[normalem]{ulem} 
\usepackage{url}
\usepackage{color,soul}
\usepackage[caption=false]{subfig}
\usepackage[T1]{fontenc}
\usepackage[utf8]{inputenc}
\usepackage{microtype}

\newcommand{\ang}[1]{\langle{#1}\rangle}

\newcommand{\chieff}{\chi_\mathrm{eff}}
\newcommand{\pd}{\partial}

\newcommand{\dual}{\,{}^*\!} 
\newcommand{\bayeswave}{\texttt{BayesWave}~}
\newcommand{\bilby}{\texttt{Bilby}~}

\newcommand{\phenom}{\texttt{IMRPhenomXPHM}~}

\newcommand{\Flatiron}{\affiliation{Center for Computational Astrophysics, Flatiron Institute, 162 5th Ave, New York, NY 10010, USA}}
\newcommand{\StonyBrook}{\affiliation{Department of Physics and Astronomy, Stony Brook University, Stony Brook NY 11794, USA}}

\newcommand{\CIT}{\affiliation{Department of Physics, California Institute of Technology, Pasadena, California 91125, USA}}
\newcommand{\CITLab}{\affiliation{LIGO Laboratory, California Institute of Technology, Pasadena, California 91125, USA}}
\newcommand{\PCC}{\affiliation{Department of Physics, Pasadena City College, Pasadena, California 91106, USA}}

\begin{document}

\title{Gravitational wave inference on a numerical-relativity simulation of a black hole merger beyond general relativity}

\author{Maria Okounkova}
\email{mokounkova@pasadena.edu}
\Flatiron \PCC

\author{Maximiliano Isi}
\email{misi@flatironinstitute.org}
\Flatiron

\author{Katerina Chatziioannou}
\email{kchatziioannou@caltech.edu}
\CIT \CITLab 

\author{Will M. Farr}
\email{wfarr@flatironinstitute.org}
\StonyBrook{} \Flatiron{}

\date{\today}

\begin{abstract}
We apply common gravitational wave inference procedures on binary black hole merger waveforms beyond general relativity. We consider dynamical Chern-Simons gravity, a modified theory of gravity with origins in string theory and loop quantum gravity. This theory introduces an additional parameter $\ell$, corresponding to the length-scale below which beyond-general-relativity effects become important.
We simulate data based on numerical relativity waveforms produced under an approximation to this theory, which differ from those of general relativity in the strongly nonlinear merger regime.
We consider a system with parameters similar to GW150914 with different values of $\ell$ and signal-to-noise ratios. We perform two analyses of the simulated data. The first is a template-based analysis that uses waveforms derived under general relativity and allows us to identify degeneracies between the two waveform morphologies.
The second is a morphology-independent analysis based on {\tt BayesWave} that does not assume that the signal is consistent with general relativity. The {\tt BayesWave} analysis faithfully reconstructs the simulated signals. However, waveform models derived under general relativity are unable to fully mimic the simulated modified-gravity signals and such a deviation would be identifiable with existing inference tools. 
Depending on the magnitude of the deviation, we find that the templated analysis can under perform the morphology-independent analysis in fully recovering simulated beyond-GR waveforms even for achievable signal-to-noise ratios $\gtrsim 20{-}30$.
\end{abstract}

\maketitle

\section{Introduction}
\label{sec:introduction}

Although general relativity (GR) has passed all precision tests to date, it is expected to break down at some scale where gravity is reconciled with quantum mechanics through a beyond-GR theory of gravity. Moreover, considering possible modifications to GR and testing for them can help shed light on the theory itself. 
Binary black hole (BBH) mergers can probe gravity at its most extreme: in the strong-field, highly non-linear, highly-dynamical regime, and are therefore an exceptional laboratory for detecting beyond-GR physics~\cite{Baker:2014zba, Will:2014kxa, Berti:2015itd}. Thus far, LIGO-Virgo detections of gravitational wave (GW) signals from BBH systems have been shown to be consistent with GR~\cite{TheLIGOScientific:2016src,Yunes:2016jcc, LIGOScientific:2019fpa, LIGOScientific:2020tif, LIGOScientific:2021sio}.

Tests of GR with GWs can be broadly divided into two categories: unmodeled and modeled. 
The first kind, such as residuals tests, looks for generic consistency between GR waveform templates and the observed data, without a specific model for GR deviations~\cite{LIGOScientific:2016lio,LIGOScientific:2019fpa,LIGOScientific:2020tif,Ghonge:2020suv,LIGOScientific:2021sio,Ghosh:2016qgn,Ghosh:2017gfp, Maselli:2019mjd}.
The second seeks inspiration from beyond-GR dynamics to look for GW properties beyond GR, or introduce phenomenological deviations to the observed signals, such as the parametrized post-Einsteinian framework or ringdown tests \cite{Yunes:2009ke,Chatziioannou:2012rf,Agathos:2013upa,PhysRevD.90.064009,Isi:2019aib,Carullo:2019flw,Brito:2018rfr}.
For either kind of test, it is beneficial to have concrete examples of plausible waveforms beyond GR; unfortunately, such examples have typically been restricted to regimes that are tractable analytically, namely the binary inspiral or ringdown.
Recently, however, there has been progress toward numerical relativity simulations of BBHs in theories beyond GR~\cite{Okounkova:2019dfo, Okounkova:2019zjf, Okounkova:2020rqw, East:2020hgw, Witek:2018dmd, Cayuso:2017iqc, Cayuso:2020lca, Allwright:2018rut,Ripley:2022cdh,Silva:2020omi,Witek:2020uzz,Hirschmann:2017psw,Berti:2013gfa,Healy:2011ef}. Even though some of these simulations solve the underlying field equations only approximately, they can still serve as a qualitative example of how beyond-GR dynamics might modify the strongly nonlinear merger BH regime.

Our aim in this study is to explore how existing GW inference tools can recover signals whose merger does not follow GR. We use the beyond-GR waveforms computed through numerical relativity to simulate LIGO~\cite{LIGOScientific:2014pky} and Virgo~\cite{VIRGO:2014yos} data and study the behavior of two complementary inference analyses.  The first, based on {\tt BayesWave}~\cite{Cornish:2014kda,Littenberg:2014oda,Cornish:2020dwh}, follows a morphology-independent approach and the second, based on {\tt Bilby}~\cite{Ashton:2018jfp,Romero-Shaw:2020owr}, assumes that the signal is consistent with existing waveform templates derived within GR.
For this study, we do not consider parameterized tests of GR deviations since those are typically anchored to physical models in the inspiral and ringdown regimes, not the merger.

The {\tt BayesWave} morphology-independent analysis does not assume that the signal follows the time-frequency evolution prescribed by GR, but it requires that the signal be coherent across the network and that it travel at the speed of light. Though it is possible to relax them in targeted analyses, we make two additional assumptions: (i) the signal contains solely tensor polarization modes~\cite{Chatziioannou:2021mij}, and (ii) it is elliptically polarized~\cite{Cornish:2020dwh}. With these restrictions, the analysis is expected to recover and faithfully reconstruct beyond-GR signals regardless of their precise time-frequency evolution.

\begin{figure}
  \includegraphics[width=0.9\columnwidth]{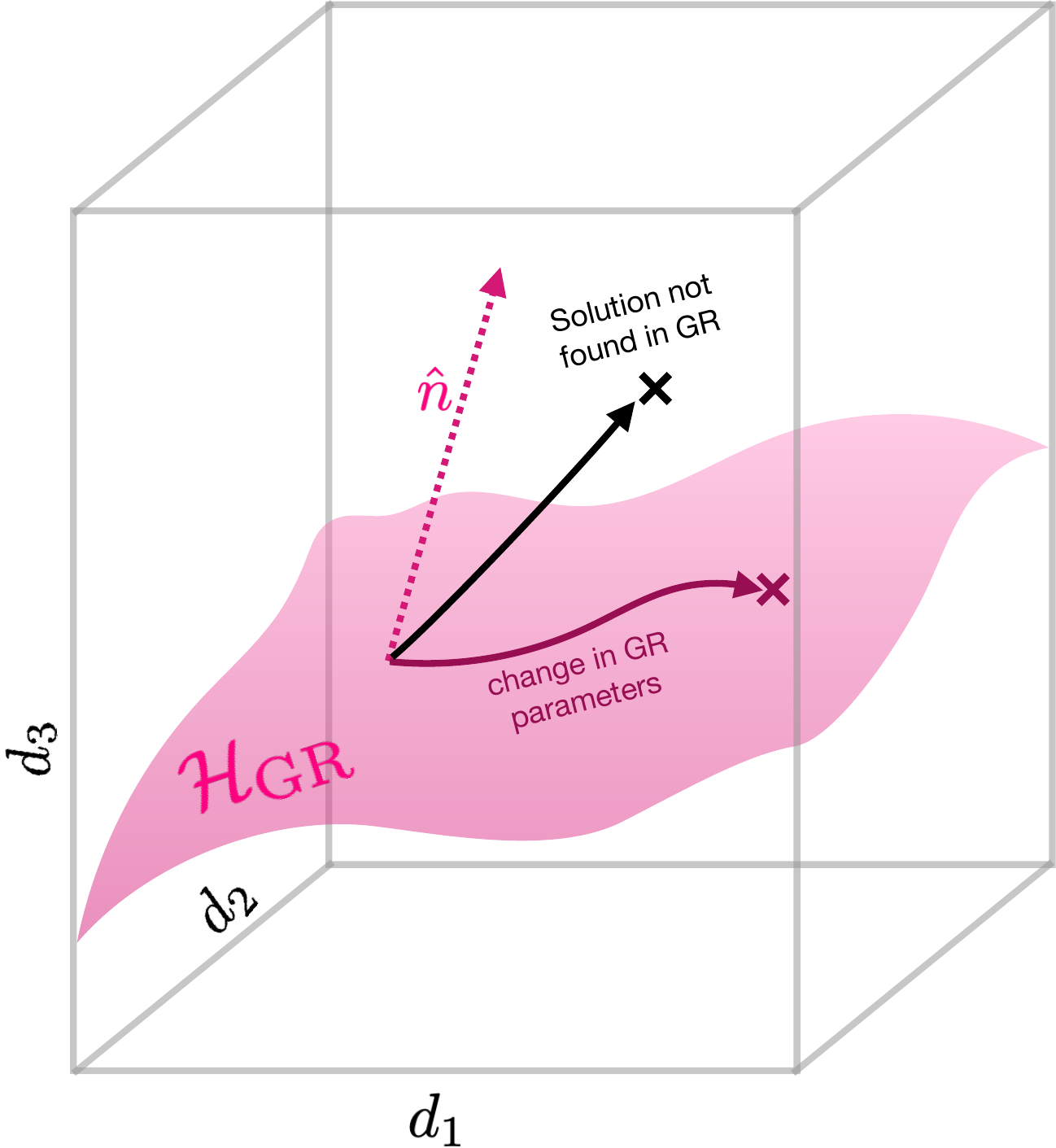}
  \caption{Illustration of degeneracy between GR and beyond-GR theories. BBH GW data lives in a space $\mathcal{D}$ (represented by the gray cube), where each dimension (here only $d_1$, $d_2$, and $d_3$ are shown) corresponds to the value of the waveform at discrete frequencies. The set of GR vacuum BBH gravitational waveforms forms a hypersurface $\mathcal{H}_\mathrm{GR}$ (represented by the pink surface) in $\mathcal{D}$. In other words, the gravitational waveform for any BBH merger in GR will be found on $\mathcal{H}_\mathrm{GR}$.
  Changing the parameters of the GR BBH system, such as the masses and the spins, corresponds to movement along $\mathcal{H}_\mathrm{GR}$ (maroon curve) from one GR solution to another. 
  Let us consider a point in $\mathcal{H}_\mathrm{GR}$, and let $\hat{n}$ (dashed pink line) be the normal to $\mathcal{H}_\mathrm{GR}$ at that point. Then introduce a beyond-GR modification to the gravitational waveform: if beyond-GR and GR are morphologically non-degenerate, this modification will include a component along $\hat{n}$ (and can include a component along $\mathcal{H}_\mathrm{GR}$). If, however, there is degeneracy between a beyond-GR theory and GR, then beyond-GR modifications will only include movement along $\mathcal{H}_\mathrm{GR}$, so the non-GR effect can be accommodated by a change in GR parameters, amounting to stealth bias.}
  \label{fig:BeyondGRSpace}
\end{figure}

The {\tt Bilby} template-based analysis, meanwhile, uses quasicircular BBH waveforms computed within vacuum GR with no extra physics, to model signals and estimate their physical source parameters within the GR BBH parameter space.

This analysis can illuminate degeneracies between GR and beyond-GR theories, as well as potential biases in recovered GR BBH parameters, as illustrated in Fig.~\ref{fig:BeyondGRSpace}. If beyond-GR modifications are degenerate with GR, then the template-based analysis will be able to fully model the signal, albeit with a potential \emph{stealth} bias on the BBH parameters~\cite{Vitale:2013bma, Vallisneri:2013rc}. If, however, the GR and beyond-GR modifications are non-degenerate, then the template-based analysis will leave a residual signal behind. 

In this study, we focus on BBH mergers in dynamical Chern-Simons (dCS) gravity. dCS is a beyond-GR theory that adds a pseudo-scalar field coupled to spacetime curvature to the Einstein-Hilbert action, and has origins in string theory and loop quantum gravity~\cite{Alexander:2009tp, Green:1984sg, Taveras:2008yf, Mercuri:2009zt}. This coupling is governed by a parameter $\ell$, which has dimensions of length and corresponds to the scale below which beyond-GR effects become important; GR is recovered for $\ell = 0$. We use the waveforms from~\cite{Okounkova:2019zjf} to simulate data for a system consistent with GW150914~\cite{Abbott:2016blz}, for a variety of dCS coupling constants $\ell$ (including $\ell = 0$) and signal-to-noise ratios (SNRs).

We find that the morphology-independent {\tt BayesWave} analysis can reconstruct the injected beyond-GR signals for all values of $\ell$, with the fidelity of the reconstruction improving with the SNR of the simulated signal. Meanwhile, the GR-templated {\tt Bilby} analysis \textit{cannot} fully reconstruct the injected beyond-GR signals and recovers biased source parameters. This shows that the effects introduced in the merger phase of this beyond-GR waveform~\cite{Okounkova:2019zjf} are non-degenerate with GR. At sufficiently high SNR ($\gtrsim 20-30$, depending on the value of $\ell$), existing inference tools can identify discrepancies between the observed signal and the GR expectation. Fully interpreting such a discrepancy, however, would require additional consideration and careful studies of waveform systematics even within GR~\cite{Purrer:2019jcp}. 

This paper is organized as follows: In Sec.~\ref{sec:waveforms}, we detail the beyond-GR numerical relativity merger waveforms we use. In Sec.~\ref{sec:methods}, we describe the methods for the morphology-independent and the template-based analysis, while in Sec.~\ref{sec:results} we describe the results of the two methods and how they compare. We conclude in Sec.~\ref{sec:discussion}.

The code and documentation for performing and reproducing this analysis, including the numerical relativity waveforms used in this study, are available at~\cite{release}.

\section{Beyond-GR Gravitational Waveforms and Simulated data}
\label{sec:waveforms}

We consider dCS~\cite{Alexander:2009tp}, a beyond-GR theory that treats the Einstein-Hilbert action of GR as the first term in a higher-order expansion in spacetime curvature. The higher-order curvature terms, though classical, are inspired by quantum gravity corrections~\cite{Green:1984sg, Taveras:2008yf, Mercuri:2009zt}. This expansion is governed by a parameter $\ell$, which corresponds to the length-scale below which beyond-GR gravity effects become important. The dCS action takes the form (setting the speed of light $c = 1$ throughout)
\begin{align}
\label{eq:dCSAction}
S \equiv \int d^4 x \sqrt{-g} \left( \frac{R}{16 \pi G}  - \frac{\ell^2}{16 \sqrt{2 \pi G}}  \vartheta \dual RR - \frac{1}{2} (\pd \vartheta)^2  \right) ,
\end{align}
where the first term is the Einstein-Hilbert action of GR and the second term introduces a quadratic curvature effect through the Pontryagin density, $\dual RR \equiv \dual R^{abcd} R_{abcd}$, with coupling $\ell^2$.\footnote{$\dual R^{abcd} R_{abcd}$ refers to the dual of a the Riemann tensor contracted with itself, which can be expressed using the fully antisymmetric Levi-Civita tensor $\epsilon_{abcd}$ as $\dual R^{abcd} R_{abcd} = \frac{1}{2} \epsilon^{abef} R_{ef}{}^{cd} R_{abcd}$.}
In order to make the theory dynamical, we couple the quadratic curvature term to an axionic scalar field $\vartheta$, where the final term in the action corresponds to the kinetic term for the scalar field. 
 
In order to ensure well-posedness of the dCS evolution equations~\cite{Delsate:2014hba}, the corresponding numerical relativity simulations are performed in an \textit{order-reduction scheme}~\cite{Okounkova:2017yby}, in which the spacetime metric and the dCS scalar field are perturbed around GR. Because the coupling in Eq.~\eqref{eq:dCSAction} is $\ell^2$, each order $n$ in the expansion will take the power $\ell^{2n}$. Zeroth order ($\ell^0$) corresponds to a GR background spacetime. At first order ($\ell^2$), the curvature of the GR background sources the leading-order scalar field. At second order ($\ell^4$), the curvature of the GR background and the leading-order scalar field source the leading-order correction to the spacetime metric. This scheme gives us access to the background GR strain waveform $h_\mathrm{GR}$ and the leading order dCS correction to the strain, $\ell^4 \delta h_\mathrm{dCS}$. The total leading-order corrected dCS waveform is the sum of the two using the dCS coupling constant as
\begin{align}
\label{eq:hmod}
    h = h_\mathrm{GR} + (\ell/GM)^4 \delta h_\mathrm{dCS}\,,
\end{align}
where $M$ the total mass of the binary. The larger the value of $\ell$, the larger the beyond-GR effects. The order-reduction scheme allows us to evaluate $h_\mathrm{GR}$ and $h_\mathrm{dCS}$ once given some system parameters, and then generate a beyond-GR waveform for any value of the coupling constant. However, due to the perturbative nature of the scheme, there is an instantaneous regime of validity (see~\cite{Okounkova:2017yby,Okounkova:2019dfo,Okounkova:2019zjf} for technical details), that limits the allowed values of $\ell/GM$. The resulting waveform is not an exact solution to the dCS field equations, but it can still be used as an example of beyond-GR dynamics in a data analysis setting.

\subsection{Simulated data parameters}
\label{sec:simulation_parameters}

For this study, we use the waveform from~\cite{Okounkova:2019zjf} with parameters consistent with GW150914~\cite{TheLIGOScientific:2016wfe, Kumar:2018hml, Abbott:2016blz, Lovelace:2016uwp} in the $\ell = 0$ (GR) case. We have chosen such a system because GW150914 is well-studied, including with many tests of GR~\cite{Yunes:2016jcc, LIGOScientific:2016lio,Isi:2019aib}. We choose a total mass of $M \simeq 68 M_\odot$, consistent with GW150914~\cite{TheLIGOScientific:2016wfe}. This choice further ensures that most of the signal observed by LIGO-Virgo is near the merger phase where the dCS modification is the strongest.
As discussed in~\cite{Okounkova:2019zjf}, the order-reduction scheme leads to \textit{secular} growth during the inspiral between the `full' and `perturbed' dCS solutions. To avoid this secular growth, the numerical simulations of~\cite{Okounkova:2019zjf} give a beyond-GR waveform with the dCS effects smoothly ramped-on close to merger, thus producing a combination of a GR inspiral with a dCS merger. Future work will include dCS modifications to the inspiral phase as well~\cite{GalvezGhersi:2021sxs}. 

\begin{figure}
  \includegraphics[width=\columnwidth]{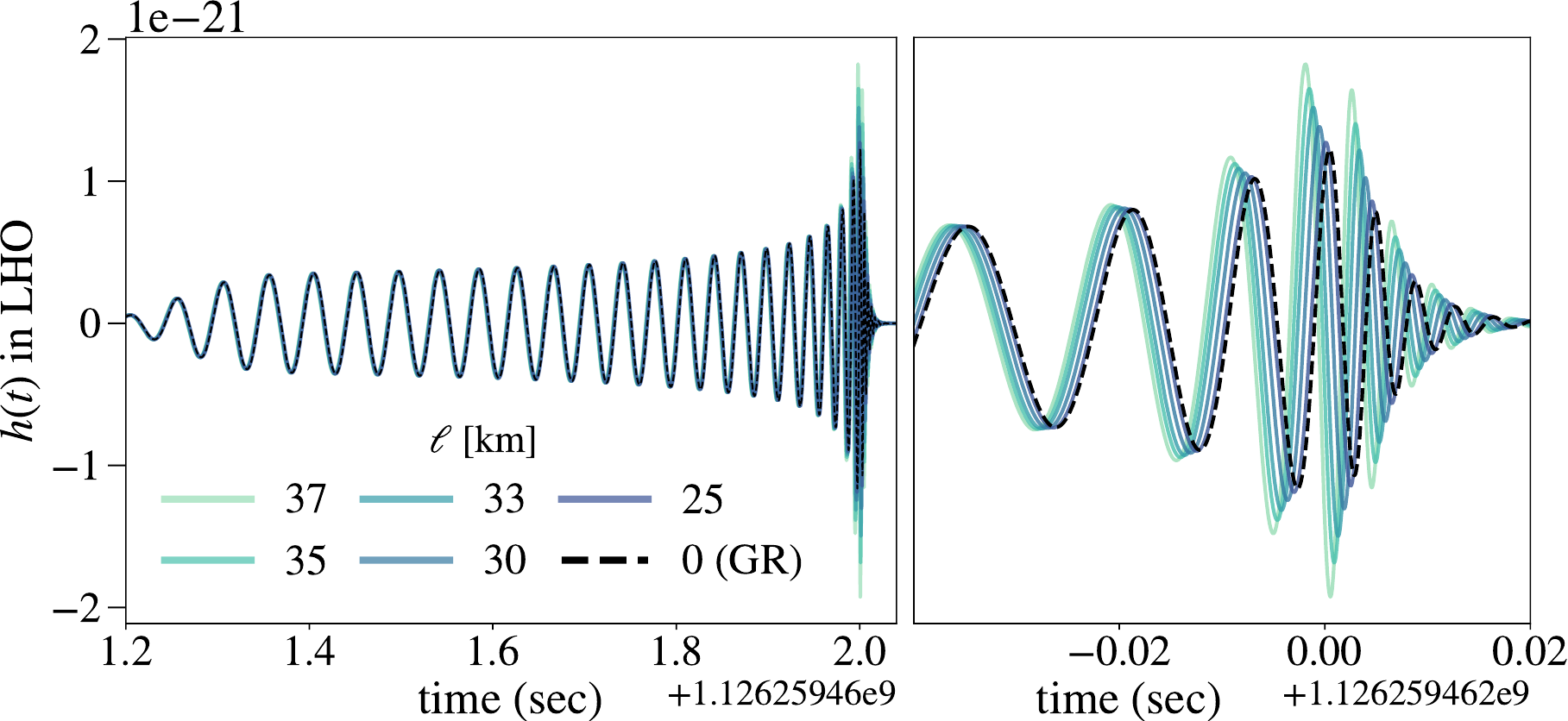}
  \caption{Gravitational strain in LHO for various values of the dCS coupling parameter $\ell$ and an injected distance of 500 Mpc. The waveforms are aligned at the start and are smoothly ramped on from zero. The black curve with $\ell = 0$ corresponds to the GR waveform, while the remaining curves show beyond-GR modified waveforms for various values $\ell$. With increasing $\ell$, the phase of the beyond-GR waveform evolves more rapidly relative to that of GR, leading to the beyond-GR waveforms peaking earlier~\cite{Okounkova:2019zjf}. The beyond-GR waveforms also have a larger amplitude at merger~\cite{Okounkova:2019zjf} and thus have increased network SNRs at the same luminosity distance.}
  \label{fig:dCSWaveforms}
\end{figure}

The physical parameters of the simulated system are consistent with those of the numerical relativity waveform in Fig. 1 of~\cite{Abbott:2016blz} as well as follow-up studies~\cite{Lovelace:2016uwp, Bhagwat:2017tkm, Giesler:2019uxc}. The initial dimensionless spins vectors are $\vec{\chi}_1 = 0.330 \hat{z}$ and $\vec{\chi}_2 = -0.440 \hat{z}$, aligned and anti-aligned with the orbital angular momentum respectively, leading to no spin-precession effects. The ratio of the component masses is $q = 0.819$, and the remnant BH has final Christodoulou mass $0.9525$ in units of the initial mass of the system, and dimensionless spin $0.692$ purely in the $\hat{z}$ direction. In vacuum numerical relativity simulations, the total mass of the system is scaled out, and thus when performing injection studies, we can introduce an arbitrary total mass. To be consistent with GW150914, we choose $M = 68 M_\odot$~\cite{TheLIGOScientific:2016wfe}. When projecting the waveform into a detector, we choose a geocenter GPS time of $1126259460$\,s, a right ascension of $1.95$ radians, a declination of $-1.27$ radians, and a binary inclination of $\pi$ radians. We show waveforms for various values of $\ell$ in Fig.~\ref{fig:dCSWaveforms}. The largest value we choose for $\ell$ corresponds to the maximum-allowed value for our order-reduction-schemes to be within the instantaneous regime of validity (cf.~\cite{Okounkova:2019zjf}). We choose spacing of $\ell$ values roughly even in $\ell^4$, the order at which beyond-GR affects appear in the waveform. 

We use these waveforms to simulate data observed with LIGO Hanford (LHO), LIGO Livingston (LLO), and Virgo using the infrastructure of~\cite{Schmidt:2017btt} and assuming a zero-noise realization. We vary the strength of the signal by changing the luminosity distance of the system, computing the SNR using the Advanced LIGO \textit{design} sensitivity~\cite{LIGOPSD} curve for LHO, LLO, and Virgo. Current BBH observations have typical SNRs $~\sim 10-25$~\cite{LIGOScientific:2021djp}, with GW150914 having a network SNR of 24 in two LIGO detectors~\cite{Abbott:2016blz} with the O1 sensitivity. Next-generation GW detectors will detect BBH mergers with SNRs reaching and exceeding $\sim100$~\cite{Sathyaprakash:2012jk, Evans:2016mbw}. We thus consider network SNRs in the range 25-125 in order to span a wide range of detector capabilities. 

\section{Methods}
\label{sec:methods}

Given the simulated data described in Sec.~\ref{sec:waveforms}, we consider both a flexible morphology-independent and a GR-template based analysis and describe them in this section. Both analyses target the same data and a frequency band of $(f_\mathrm{low},f_\mathrm{high})=(25, 1024)$ Hz. The low frequency limit is determined by the finite length of the numerical relativity simulation, while the high frequency limit is chosen such that the analysis includes the binary merger and ringdown.

We quantify how well the two analyses reconstruct the simulated signal through the overlap between different waveforms, see e.g.,~\cite{Apostolatos:1995pj,Ghonge:2020suv}. For two waveforms $A$ and $B$ the overlap is defined as
\begin{align}
\label{eq:overlap}
    \mathcal{O}_{\mathrm{A},\mathrm{B}} \equiv \frac{\ang{h_\mathrm{A}, h_\mathrm{B}}_N}{\sqrt{\ang{h_\mathrm{A}, h_\mathrm{A}}_N \ang{h_\mathrm{B}, h_\mathrm{B}}_N}}\,.
\end{align}
where
\begin{align}
    \ang{h_\mathrm{A}, h_\mathrm{B}}_N &= \sum_{i}^N \ang{h_\mathrm{A}^i, h_\mathrm{B}^i}\,, \\
    \ang{h_\mathrm{A}^i, h_\mathrm{B}^i} &= 4 \mathrm{Re} \int_{f_\mathrm{low}}^{f_\mathrm{high}}\frac{\tilde{h}_\mathrm{A}^i(f) \tilde{h}_\mathrm{B}^{i*}(f)}{S_n^i (f)} df\,.\label{eq:inner_product}
\end{align}
Here $h_\mathrm{A}^i$ and $h_\mathrm{B}^i$ are the two target waveforms as seen in detector $i$, $\tilde{h}$ denotes the frequency-domain waveform, $S_n^i(f)$ is the noise power spectral density (PSD) of detector $i$, and $N$ is the total number of detectors. The optimal network SNR is $\sqrt{\ang{h, h}}$.
For waveforms that agree perfectly, the overlap is 1. We also define the \textit{mismatch} between two waveforms as 
\begin{align}
\label{eq:mismatch}
    \Delta_{\mathrm{A},B} \equiv 1 - \mathcal{O}_{\mathrm{A},\mathrm{B}}\,.
\end{align}

\subsection{Morphology-independent analysis}
\label{sec:bayeswave_methods}

We use \bayeswave~\cite{Cornish:2014kda, Littenberg:2014oda, Cornish:2020dwh} to perform a morphology-independent analysis that does not impose that the signal be consistent with GR dynamics. \bayeswave models the data through a sum of sine-Gaussian wavelets~\cite{morlet}, requiring only that the signal be coherent across the detector network, travel at the speed of light between detectors, contain only tensor polarizations in an elliptical configuration, and do not disperse while traveling between the detectors. These assumptions are satisfied by our simulated data. Indeed, in dCS, GWs travel at the speed of light between detectors and contain only tensor polarizations~\cite{Alexander:2009tp}. Moreover, the numerical relativity simulation we use in thus study (cf. Sec.~\ref{sec:simulation_parameters}) does not exhibit spin-precession that could break the elliptical polarization assumption. 
Besides these, \bayeswave makes no further assumption about the time-frequency content of the signal being consistent with GR. 

Details about the wavelet model and the sampler implementation, settings, and priors are provided in~\cite{Cornish:2014kda, Littenberg:2014oda,Cornish:2020dwh,Wijngaarden:2022sah}. Here we specifically use the ``signal model" with a fixed PSD,
and sample the multi-dimensional posterior for the number of the sine-Gaussian wavelets, their parameters, and the signal extrinsic parameters using a collection of (Reversible Jump) Markov Chain Monte Carlo~\cite{10.1093/biomet/82.4.711} samplers.
Under the default configuration and priors, \bayeswave is more sensitive to BBH signals and times close to merger, where the GW amplitude peaks~\cite{Littenberg:2015kpb,2016PhRvD..93b2002K,Ghonge:2020suv}.

\subsection{Template-based analysis}
\label{sec:bilby_methods}

For the template-based analysis with GR waveforms, we use the parameter estimation software library \bilby~\cite{Ashton:2018jfp}, which has been used in parameter estimation studies of the LIGO-Virgo transient catalogs~\cite{Romero-Shaw:2020owr, LIGOScientific:2020ibl, LIGOScientific:2021djp}. \bilby uses nested sampling~\cite{skilling, Veitch:2009hd} to sample the posterior distribution for the BBH parameters. 
The BBH parameter space within GR includes the component masses $m_1$ and $m_2$, the BH spin vectors $\vec{\chi_1}, \vec{\chi_2}$, the sky location and luminosity distance of the source, the inclination angle of the total angular momentum of the binary, the polarization angle, as well as a time and phase. We use standard priors and settings~\cite{Romero-Shaw:2020owr}. For the template we use the waveform model \phenom~\cite{Pratten:2020ceb} as implemented in {\tt LALSuite}~\cite{lalsuite} due to its computational efficiency. This model includes effects of spin-precession and higher-order multiple moments and was used in the LIGO-Virgo GWTC-3 analysis~\cite{LIGOScientific:2021djp}. We have verified that the mismatch between our numerical relativity GR waveform and the \phenom waveform with the corresponding set of GR parameters is lower than the mismatch between our numerical dCS waveform and GR waveform (at the $10^{-4}$ level), showing that we can resolve dCS effects.

Rather than the component spins, we express results through common spin \textit{combinations}. The effective inspiral spin $\chieff$~\cite{Ajith:2009bn, Santamaria:2010yb} is the mass-weighted ratio sum of the spin components along the Newtonian orbital angular momentum $\hat{L}$
\begin{align}
    \label{eq:chi_eff}
    \chieff = \frac{ \vec{\chi_1} \cdot \hat{L}+ q \vec{\chi_2} \cdot \hat{L}}{1+q} \in [-1,1]\,,
\end{align}
and it is approximately conserved under spin-precession~\cite{Racine:2008qv}. Complementary to $\chieff$ is the effective precession spin $\chi_p$, which measures the mass-weighted in-plane spin component and characterizes spin precession~\cite{Hannam:2013oca, Schmidt:2014iyl}
\begin{align}
\label{eq:chi_p}
    \chi_p = \mathrm{max} \left\{ \chi_{1, \perp}, \frac{q(4q + 3)}{4 + 3q} \chi_{2, \perp}  \right\} \in [0,1]\,,
\end{align}
where $\chi_{i, \perp}$ is the component of spin perpendicular to the direction of the Newtonian orbital angular momentum. A vanishing $\chi_p$ corresponds to a system with no spin-precession, and hence spins aligned with $\hat{L}$. 

\section{Analysis Results}
\label{sec:results}

\begin{figure*}
  \includegraphics[width=\textwidth]{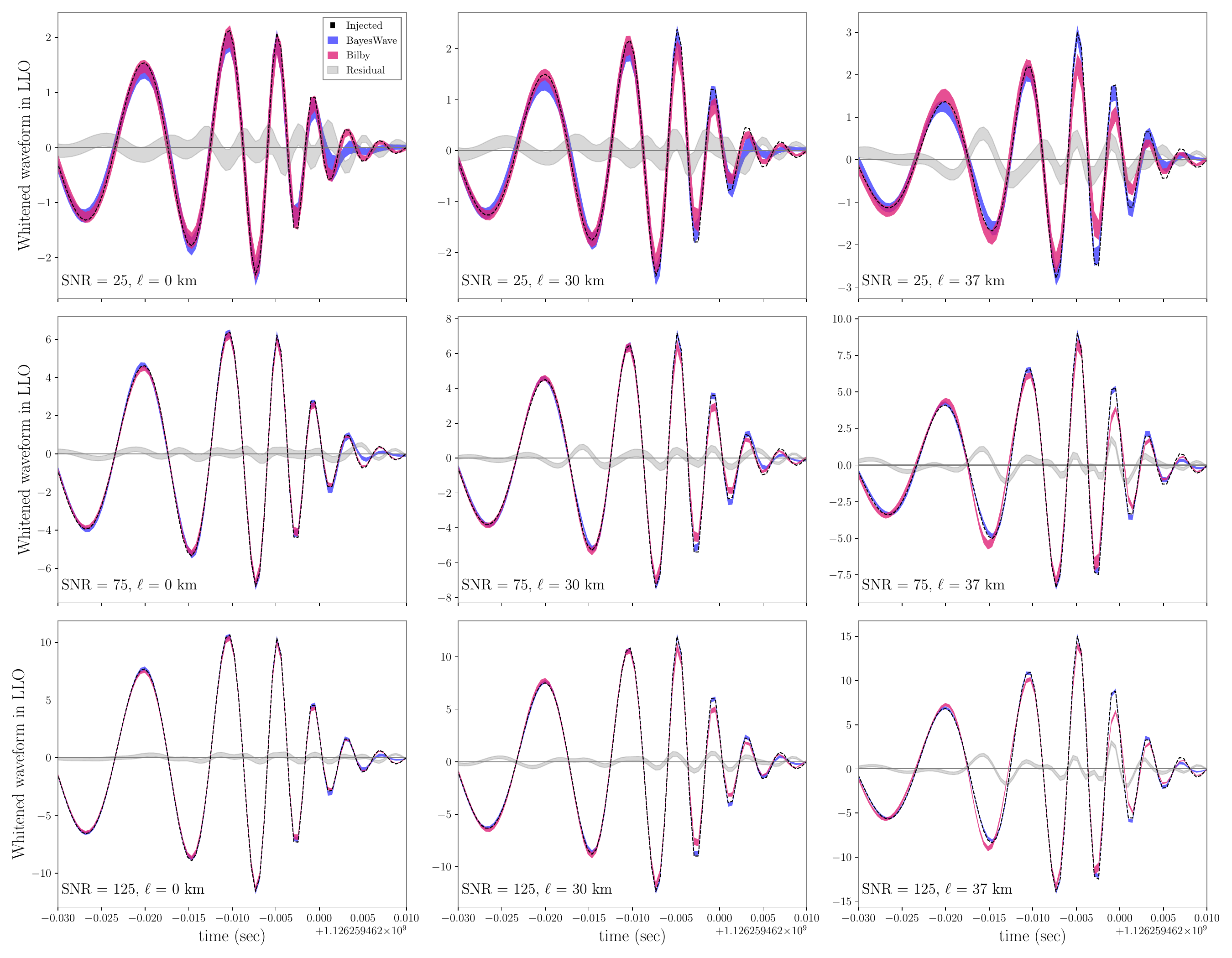}
  \caption{Time-domain whitened waveforms in LLO for different values of $\ell$ (left to right) and injected network SNR (top to bottom). The dashed black curve corresponds to the simulated signal, while blue and red shaded regions denote the 90\% credible intervals for the \bayeswave and \bilby reconstructions respectively. The \bayeswave reconstruction fully overlaps with the injected waveform for all $\ell$, with the agreement improving with SNR. The \bilby reconstruction, on the other hand, is inconsistent with the injected signal and the \bayeswave reconstruction for sufficiently high $\ell$ and SNR. We further demonstrate this by including a gray band showing the residual between the \bayeswave and \bilby results in each panel. We compute the residual interval point-wise in time by subtracting the maximum likelihood \bilby waveform from the bounds on the 90\% credible \bayeswave interval.  In the $\ell = 0$ case, the residual is consistent with zero at each time, but as we increase the value of $\ell$, the residual disagrees with zero at certain times, especially near the merger (note that the scale of the ordinate axis varies for each plot). We quantitatively assess this disagreement between the \bayeswave and \bilby results in Fig.~\ref{fig:BWLIMismatch}.
  }
  \label{fig:Bayeswave}
\end{figure*}

Figure~\ref{fig:Bayeswave} shows the whitened time-domain simulated signal and 90\%-credible intervals for the \bayeswave and \bilby reconstructions in LHO. In the case of GR ($\ell = 0$), both methods return reconstructions that agree with the simulated data, and increasingly so for higher SNR signals. While this is the expected behavior for {\tt BayesWave}'s morphology-independent approach, it is a nontrivial statement for the \bilby analysis. Waveform models such as \phenom are subject to systematics and are thus expected to not perfectly recover GR signals at sufficiently high SNRs. However, our results confirm that \phenom remains faithful to this GR numerical relativity simulation even at the highest SNRs we consider. Thus any unfaithfulness we find when $\ell \neq 0$ can be attributed to beyond-GR effects and not within-GR waveform systematics. 

When $\ell$ is nonzero, the \bayeswave reconstruction continues to accurately reproduce the simulated data, thus living up to the expectations for a morphology-independent analysis. However, the \bilby credible interval \textit{fails} to fully contain the simulated beyond-GR waveform, with the discrepancy increasing as $\ell$ increases. This suggests that the beyond-GR effect we consider here is not degenerate with GR, corresponding to the black-cross case from Fig.~\ref{fig:BeyondGRSpace}. For a real observed signal, we will thus be able to compare the \bayeswave and \bilby reconstructions and identify unexpected dynamics in the merger phase. 

We further quantify this through the ``residual,'' computed as the difference between the \bayeswave and \bilby reconstructions. 
Specifically, we subtract the maximum likelihood \bilby waveform from the upper and lower 90\% credible intervals of the \bayeswave reconstruction as a function of time to obtain a region for the residual. This illustrates the extent to which the \bilby best-fit waveform is contained in the \bayeswave interval at that specific time.
In the GR case this residual is consistent with zero where the signal is strong. As $\ell$ increases the residual becomes inconsistent with zero during the merger phase, with smaller uncertainty as the SNR increases.
This would also manifest as residual power left after subtracting the best fit \bilby reconstruction from the data, as would be measured be the residuals test formulated in~\cite{LIGOScientific:2020tif, LIGOScientific:2021sio}.
Though our analysis demonstrates that such residuals can be identified with current data analysis tools, further quantitative estimates about the SNR or the amount of deviation required depend on the exact simulated signal considered.

 \begin{figure}
  \includegraphics[width=\columnwidth]{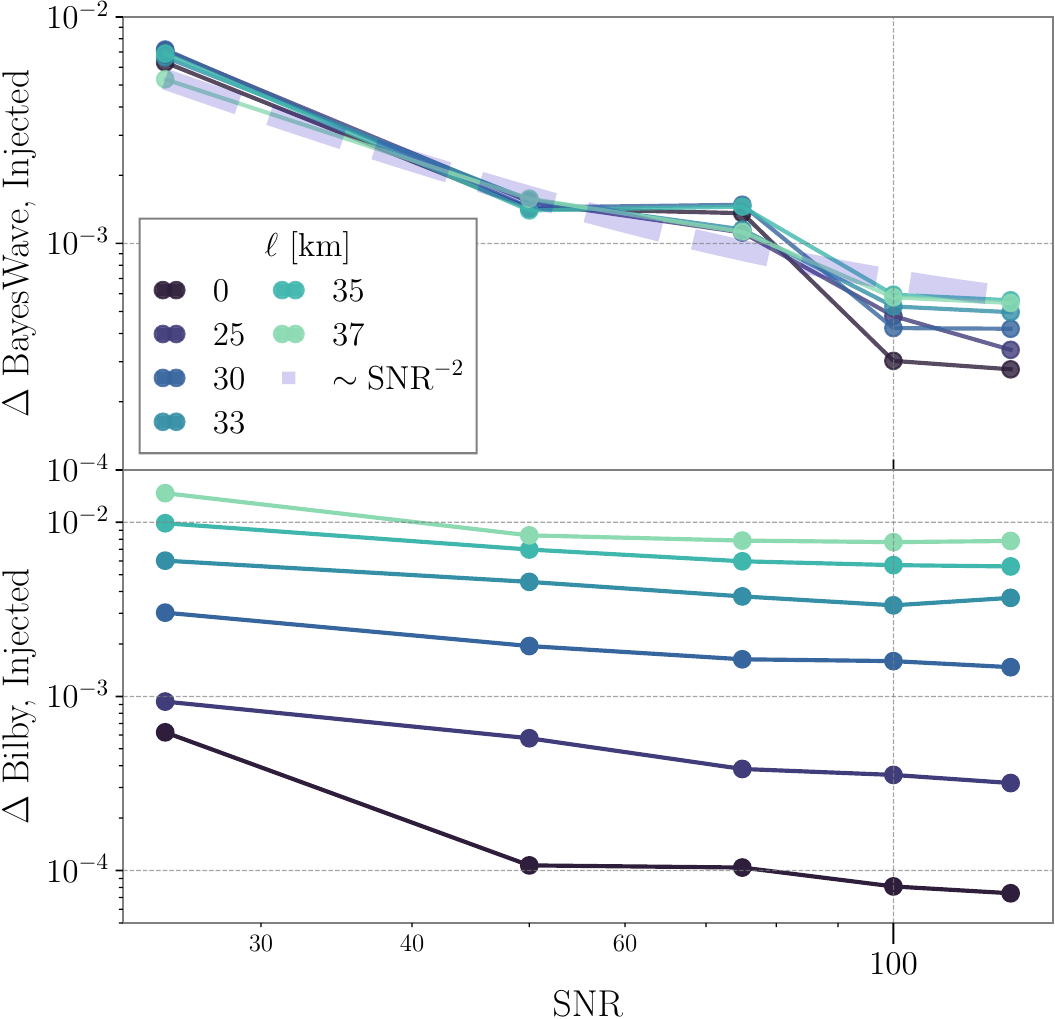}
  \caption{Mismatch between injected and reconstructed waveforms as a function of the network SNR for different values of $\ell$. The top panel corresponds to the mismatch between the injected waveforms and the median \bayeswave reconstructed waveforms, while the lower panel corresponds to the mismatch between the injected waveforms and the maximum-likelihood \bilby reconstructed waveform. The dashed purple curve corresponds to the $1/\mathrm{SNR}^2$ behavior. For all $\ell$, the \bayeswave mismatch decreases with high SNR as expected, however the \bilby mismatch deviates from this trend and plateaus with increasing $\ell$.}
  \label{fig:Mismatch}
\end{figure}

\begin{figure}
  \includegraphics[width=\columnwidth]{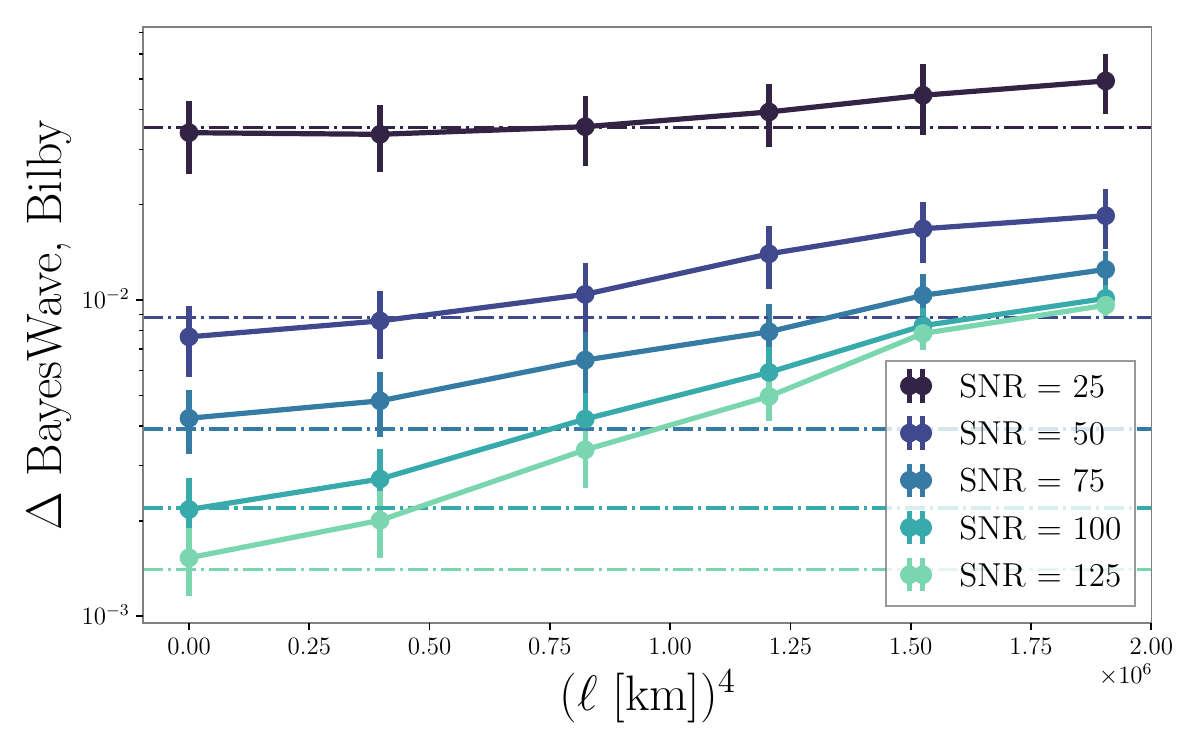}
  \caption{Mismatch between \bilby and \bayeswave reconstructed waveforms as a function of $\ell$ for a variety of injection networks SNRs. Recalling that the dCS modification comes in as a power of $\ell^4$, Eq.~\eqref{eq:hmod}, we choose to show this quantity on the horizontal axis. Each curve corresponds to a different injected network SNR. We compute mismatches between 200 \bayeswave and \bilby reconstructed waveforms sampled from their respective posterior distributions, and show the median mismatch, with the standard deviation as an errorbar. For each SNR, as the value of $\ell$ increases, the mismatch between the recovered waveforms increases. This effect persists even as we increase SNR. Each dashed line corresponds to $\mathrm{SNR}^{-2}$.
  }
  \label{fig:BWLIMismatch}
\end{figure}

This behavior is further quantified in Fig.~\ref{fig:Mismatch}, where we show the mismatch between the median \bayeswave reconstruction and the injected waveform as well as the maximum-likelihood \bilby reconstruction and the injected waveform.\footnote{The maximum-likelihood \bilby reconstructed waveform corresponds to a true GR waveform in the BBH parameter space, while the median \bilby waveform corresponds to a point-wise (in time) median taken over a set of recovered waveforms, with no guarantee that the resulting waveform corresponds to a physical GR BBH system. In the morphology-independent \bayeswave analysis, meanwhile, none of the recovered waveforms necessarily correspond to a GR BBH system, and the maximum-likelihood waveform is typically an outlier in the posterior for the number of wavelets. This is because \bayeswave is a transdimensional analysis and the maximum-likelihood waveform is typically the one with the largest number of wavelets, and thus the one that overfits the noise the most. Hence, we use the maximum-likelihood \bilby waveform and the median \bayeswave waveform when computing mismatches.} The former does not depend on the value of $\ell$ and decreases as $1/\mathrm{SNR}^2$ for high SNR as expected~\cite{Lindblom:2008cm,Baird:2012cu,Chatziioannou:2017tdw,Ghonge:2020suv}, showing that \bayeswave is able to faithfully recover both GR and beyond-GR injections. However, the latter is a strong function of $\ell$ even at SNR 25, with larger values of $\ell$ and thus large deviations from the expected GR signal leading to a larger mismatch. Additionally, the \bilby mismatch decreases less steeply with the SNR as $\ell$ increases and instead seems to plateau at large SNR. This again shows that the beyond-GR waveform cannot be faithfully reproduced with GR waveforms. Another way to demonstrate this is presented in Fig.~\ref{fig:BWLIMismatch} which shows that the mismatch between the median \bayeswave and maximum-likelihood \bilby reconstructions increases with $\ell$ for constant SNR. This figure makes only use of quantities that are available from analyses of real signals (namely, not the true injected waveform) and again demonstrates how a GR deviation can be flagged from real data.

\begin{figure}
  \includegraphics[width=\columnwidth]{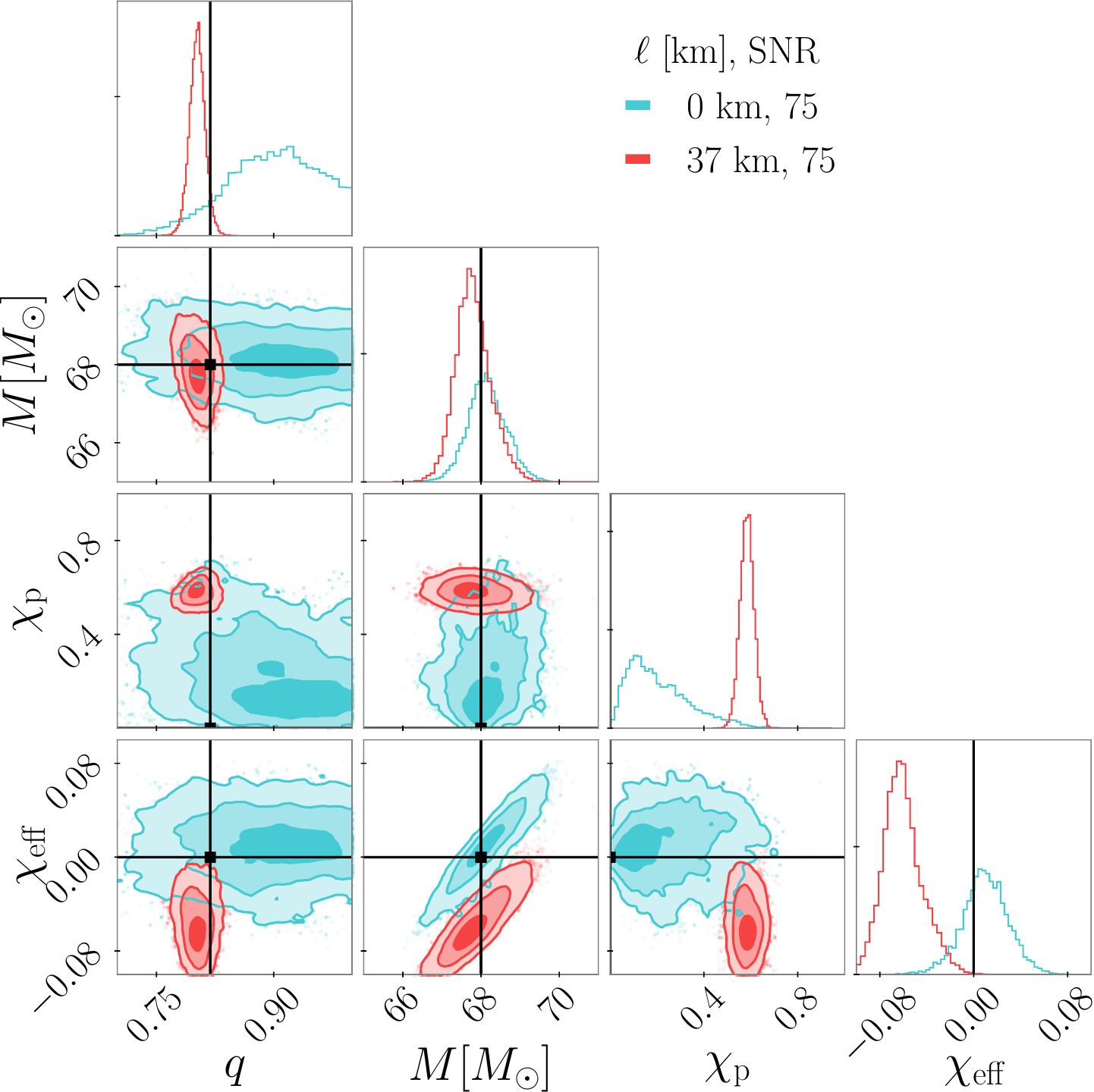}
  \caption{Marginalized one- and two-dimensional posteriors for various GR BBH parameters for a network SNR = 75 injection using a \bilby GR template-based analysis. We show $q$, the mass ratio, $M$, the detector frame total mass, $\chi_p$, the precession spin parameter, and $\chieff$, the effective spin. Blue corresponds to the posteriors for the GR ($\ell = 0$) injection, while red corresponds to the posteriors for a beyond-GR injection. The injected values are shown in black.}
  \label{fig:dCSCorner_75}
\end{figure}

\begin{figure}
  \includegraphics[width=\columnwidth]{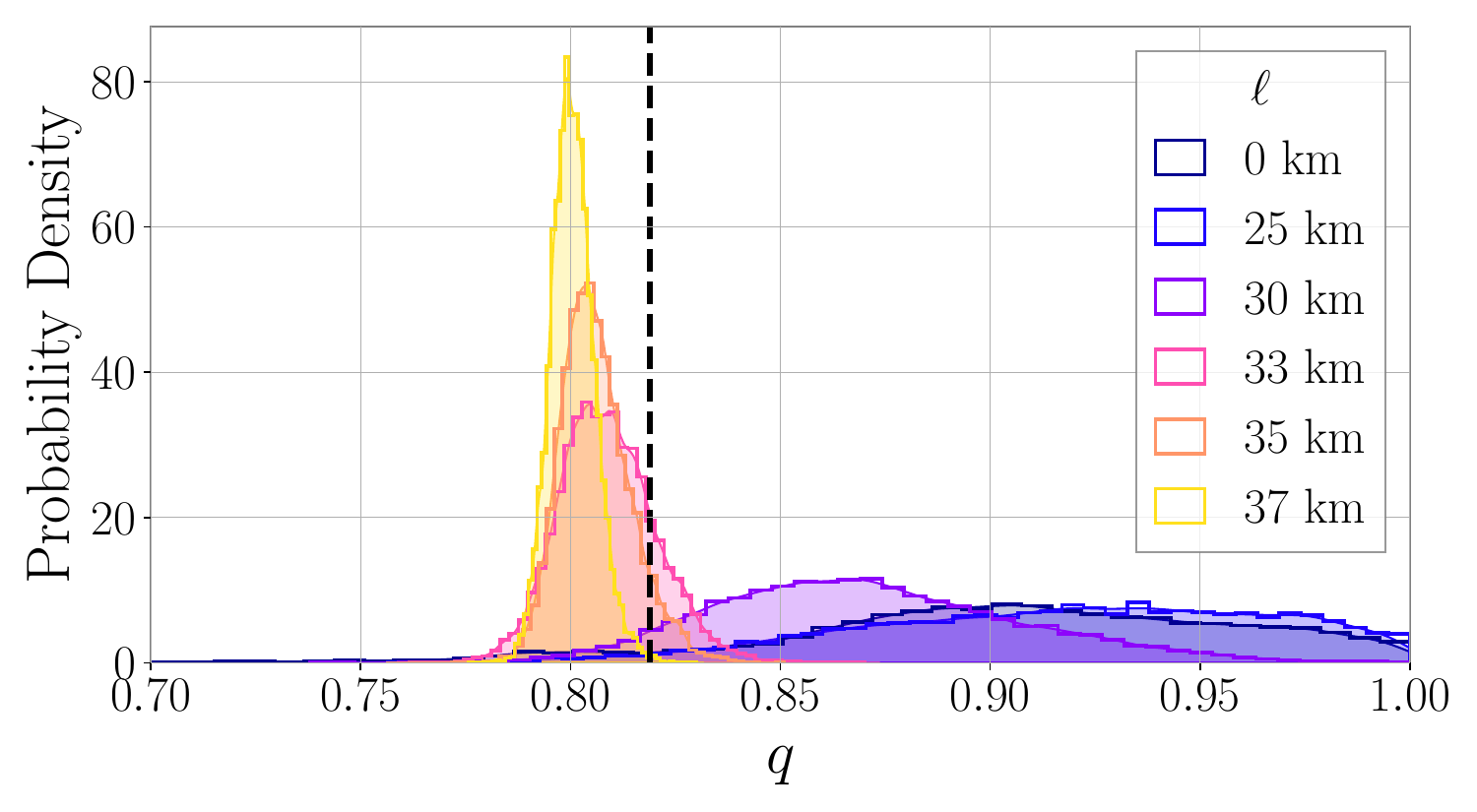}\\
  \vspace{7pt}
  \includegraphics[width=\columnwidth]{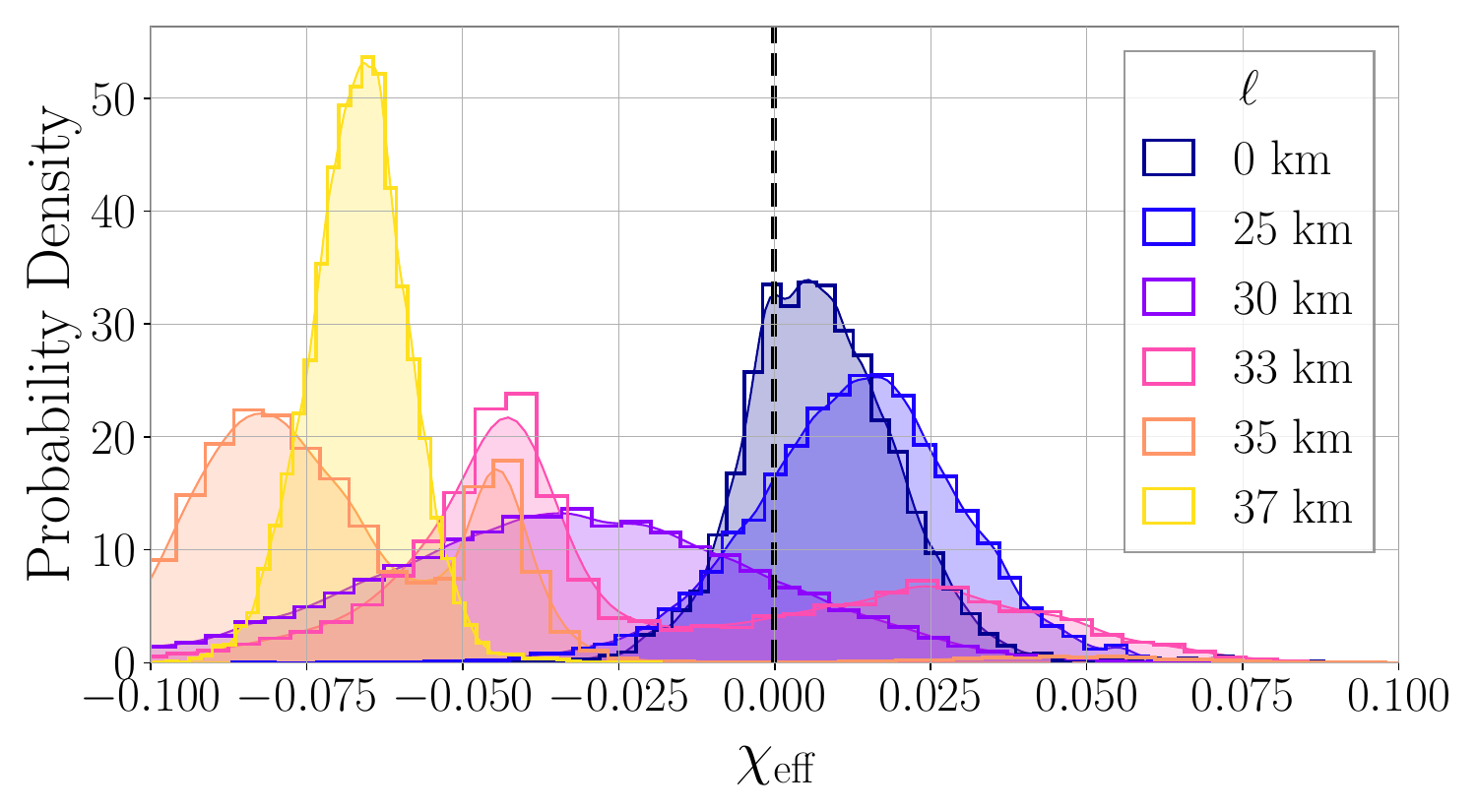} \\
  \vspace{7pt}
  \includegraphics[width=\columnwidth]{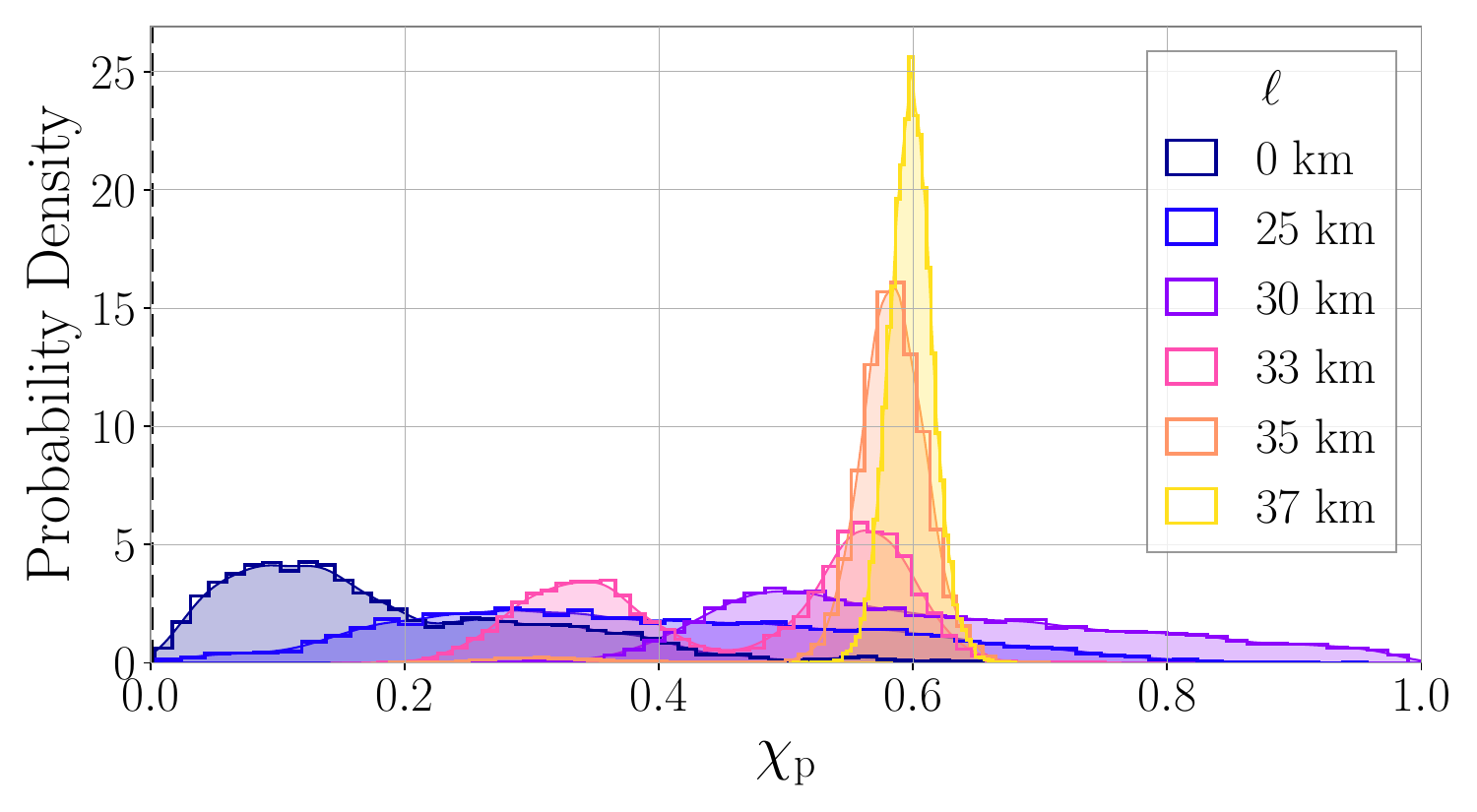}
  \caption{One-dimensional marginalized posterior for the mass ratio $q$ (top), the effective spin $\chi_\mathrm{eff}$, and the spin precession parameter $\chi_p$ (bottom), for injections with network SNR = 125. Each distribution corresponds to a different value of $\ell$, and the dashed black line corresponds to the GR value ($\chi_p = 0$ in the GR case). As $\ell$ increases the posteriors are becoming increasingly biased and peaked, a behavior discussed in Sec. \ref{sec:results}}.
  \label{fig:kde_chi_p}
\end{figure}

Figure~\ref{fig:dCSCorner_75} shows the posterior distributions for mass and spin parameters obtained with \bilby for two values of $\ell$.
When $\ell=0$, all posteriors are consistent with the injected parameters as expected. However, when $\ell \neq 0$ we recover much tighter and, in many cases, biased posteriors. The parameters with the largest posterior difference are the mass ratio $q$ and the two spin parameters. The total mass posterior is the least biased one, possibly since it is well-measured from the merger frequency. We take a closer look at the mass ratio and spins for more values of $\ell$ in Fig.~\ref{fig:kde_chi_p}. Biases in the recovered parameters are indicative of ``stealth bias,'' in which different values of GR parameters can better (though not perfectly) reproduce a beyond-GR waveform (in Fig.~\ref{fig:BeyondGRSpace} this corresponds to movement towards the beyond-GR solution parallel to $\mathcal{H}_\mathrm{GR}$). 

We can interpret the parameter biases with increasing $\ell$ as follows. The beyond-GR waveforms have roughly the same merger frequency as their GR counterparts (cf. Fig.~\ref{fig:dCSWaveforms}), but the merger is moved at earlier times and has a larger amplitude at merger. Thus, the total mass remains approximately the same to preserve the merger frequency but $\chieff$ moves to lower values, which reduces the length of the waveform due to the reduced orbital hang up effect arising from spin-orbit coupling. The mass ratio $q$ posterior remains consistent as $\ell$ changes, though it becomes increasingly peaked, a point to which we return later. Finally, a large $\chi_p$ results in large precession that can lead to a low inspiral amplitude compared to the merger amplitude. If the system precesses such that it becomes more face-on at merger compared to the inspiral, this increases the GW amplitude at merger. 

Figures~\ref{fig:dCSCorner_75} and~\ref{fig:kde_chi_p} also show that the beyond-GR posterior distributions are more sharply peaked and more tightly constrained than the GR posterior distributions. Indeed, the spread in $q$ and $\chi_p$ in the $\ell = 37$ km case is 10 times smaller than the $\ell = 0$ case. While all posteriors narrow with increased SNR, the tighter constraints on the posterior distributions in the beyond-GR case can be explained by considering the intersection of iso-likelihood contours with the signal manifold in the space of Fig.~\ref{fig:BeyondGRSpace}. The width of the likelihood distribution is controlled by the width of the intersection of such contours around the point in the manifold closest to the true signal (the maximum likelihood point); this is solely a function of the curvature of the signal manifold at that point, as would be evaluated by the Fisher matrix. When the maximum likelihood point corresponding to the beyond-GR signal lies in a region of parameter space with higher (such as low $q$) or lower (such as high $q$) curvature, then the width of the likelihood distribution in the recovered parameters decreases or increases accordingly with respect to the GR solution.

The specific point in parameter space which will maximize the likelihood will vary with the signal morphology and the preference for specific parameters (e.g., low $q$ in Fig.~\ref{fig:kde_chi_p}) is not trivial to explain, and is left for future work.

For small deviations away from GR, we would expect parameter biases to appear as linear drifts in the multi-dimensional parameter space that are proportional to $\ell^4$. This is indeed the observed behavior in $q$ and $\chi_p$ in Fig.~\ref{fig:kde_chi_p}. The inferred value of $\chieff$, however, varies with $\ell$ in a more complicated way and even displays multi-modal structure. We have verified that this $\chieff$ behavior is convergent with increasing SNR, it is therefore well-resolved and not a sampling artifact. This suggests that some simulated signals are away from the regime of small deviations at least as far as $\chieff$ is concerned.

\section{Discussion and conclusions}
\label{sec:discussion}

We performed a GW data analysis injection study using beyond-GR BBH \textit{merger} waveforms. Specifically, we focused on dCS gravity, a beyond-GR theory of gravity with motivations in string theory and loop quantum gravity~\cite{Alexander:2009tp, Green:1984sg, Taveras:2008yf, Mercuri:2009zt}.
Unlike previous work, this allows us to study how realistic analysis pipelines would respond to a full beyond-GR inspiral-merger-ringdown signal obtained by approximately solving the dynamical field equations of a concrete alternative to GR.

As detailed in Sec.~\ref{sec:waveforms}, we injected the numerical relativity merger waveforms from~\cite{Okounkova:2019zjf} that approximately solve the dCS equations through an order-reduction scheme, with parameters consistent with GW150914 into morphology-independent and template-based analysis pipelines for a variety of values of the dCS coupling constant $\ell$ and injection SNR. Though we used a specific waveform for the recovery, \phenom\cite{Pratten:2020ceb}, we found that it is faithful to the GR simulations at every SNR considered. We thus expect our main qualitative conclusions to be generic under other waveforms.

We showed that the morphology-independent analysis with \bayeswave can successfully  reconstruct beyond-GR signals for all injected values of $\ell$. This analysis makes minimal assumptions, assuming only that the same signal be detected in different interferometers, come from one sky location, travel at the speed of light between detectors, and contain only tensor polarizations. Both GR and dCS fit into this class of assumptions. Should LIGO-Virgo observe a signal with significant beyond-GR effects, a morphology-independent analysis such as \bayeswave will be able to fully reconstruct the signal. When compared to a \bilby analysis that is based on GR waveform models, comparing the two reconstructions will reveal a discrepancy and additional, unmodeled dynamics in the observed signal. Though not considered in this study, parametrized or residual tests might also be able to identify the beyond-GR effects~\cite{LIGOScientific:2016lio,LIGOScientific:2019fpa,LIGOScientific:2020tif,LIGOScientific:2021sio}, though the latter is only sensitive to large deviations~\cite{Johnson-McDaniel:2021yge}.

We return to the picture of Fig.~\ref{fig:BeyondGRSpace} to study the results of the GR-template analysis. The \bilby analysis used (all) solutions lying in $\mathcal{H}_\mathrm{GR}$ to recover the injected beyond-GR waveforms. Figure~\ref{fig:dCSCorner_75} then shows that the recovered GR system parameters from beyond-GR data differ from their injected values, hence showing motion along $\mathcal{H}_\mathrm{GR}$. However, we also showed in Fig.~\ref{fig:Bayeswave} that solutions in $\mathcal{H}_\mathrm{GR}$ did \textit{not} fully recover the beyond-GR signal, with the effect worsening for increasing $\ell$. This in turn implies that when increasing $\ell$ from zero (which lies on $\mathcal{H}_\mathrm{GR}$) there is also motion along $\hat{n}$, the normal to $\mathcal{H}_\mathrm{GR}$. In turn, the beyond-GR solution does not lie on $\mathcal{H}_\mathrm{GR}$. 

Attributing a discrepancy between the \bayeswave and \bilby reconstruction to beyond-GR effects comes, however, with additional complications. A number of effects could induce components along $\hat{n}$ besides modifications to GR, for example waveform systematics or unmodeled phenomena such as matter or orbital eccentricity. If such a discrepancy is detected in real data, extensive modeling and theory work will be required to reach a robust interpretation regarding the cause. Numerical simulations including beyond-GR dynamics could contribute to the theoretical understanding of such potential detected discrepancies. Additionally, combining constraints from multiple detections~\cite{Isi:2019asy,Isi:2022cii,Saleem:2021vph,Moore:2021eok} could help disentangle between systematics and beyond-GR physics.

\section*{Acknowledgements}

The authors are grateful for computational resources provided by the LIGO Laboratory and supported by NSF Grants PHY-0757058 and PHY-0823459.
The Flatiron Institute is supported by the Simons Foundation. 
KC was supported by NSF Grant PHY-2110111. 
Waveform injections were performed using \texttt{pycbc}~\cite{Biwer:2018osg, alex_nitz_2022_6324278}. The morphology-independent analysis was performed using the \bayeswave package~\cite{Cornish:2014kda, Littenberg:2014oda}, and the GR-template analysis was performed using the \bilby package~\cite{Ashton:2018jfp}. Post-processing was performed using the \texttt{pesummary} package~\cite{Hoy:2020vys} and the \texttt{igwn-wave-compare} library~\cite{igwn-wave-compare}. We thank Sudarshan Ghonge for help using the \texttt{igwn-wave-compare} library~\cite{igwn-wave-compare}. Corner plots were produced using the \texttt{corner} python package~\cite{corner}.

\bibliography{dCS_paper}

\end{document}